# FINITE VARIANCE SCALING ANALYSIS OF THE SOLAR NEUTRINO FLUX DATA FROM SAGE AND GALLEX-GNO DETECTORS

[1]Koushik Ghosh & [2]Probhas Raychaudhuri

[1]**Department of Mathematics**
**Dr. B. C. Roy Engineering College**
**Durgapur – 713206**
**INDIA**

[2]**Department of Applied Mathematics**
**University of Calcutta**
**92, A.P.C. Road, Calcutta-700 009**
**INDIA**

**ABSTRACT:**

The scaling behaviours of the monthly solar neutrino flux data from (1) SAGE detector during the period from 1st January 1990 to 31st December 2000; (2) SAGE detector during the period from April 1998 to December 2001; (3) GALLEX detector during the period from May 1991 to January 1997; (4) GNO detector during the period from May 1998 to December 2001; (5) GALLEX-GNO detector (combined data) from May 1991 to December 2001 and (6) average of the data from GNO and SAGE detectors during the period from May 1998 to December 2001 are analysed. In the present analysis the monthly solar neutrino flux data in each case is processed through finite variance scaling method in order to find the corresponding Hurst exponent. The study reveals that the monthly solar neutrino flux data in 1), 3), 4), 5) and 6) exhibit anti-persistent behaviour (or, negative autocorrelation) and has underlying trends while the monthly solar neutrino flux data in 2) shows the nature of a long memory process.

# INTRODUCTION:

The Hurst exponent occurs in several areas of applied mathematics, including fractals and chaos theory, long memory processes and spectral analysis. Hurst exponent estimation has been applied in areas ranging from biophysics to computer networking. Estimation of the Hurst exponent was originally developed in hydrology. However, the modern techniques for estimating the Hurst exponent come from fractal mathematics.

The mathematics and images derived from fractal exploded into the world in the 1970s and 1980s. It is difficult to think of an area of science that has not been influenced by fractal geometry. Along with providing new insight in mathematics and science, fractal geometry helped us to see the world around us in a different way. Nature is full of self-similar fractal shapes. A self-similar shape is a shape composed of a basic pattern which is repeated at multiple (or infinite) scale.

The Hurst exponent is also directly related to the 'fractal dimension', which gives a measure of the roughness of a surface. The fractal dimension has been used to measure the roughness of coastlines, for example.

Estimating the Hurst exponent for a data set provides a measure of whether the data is a pure random walk or has underlying trends. Another way to state this is that a random process with an underlying trend has some degree of autocorrelation. When the autocorrelation has a very long (or mathematically infinite) decay this kind of Gaussian process is sometimes referred to as a long memory process. Processes that we might naively assume as purely random sometimes turn out to exhibit Hurst exponent statistics for long memory processes.

The value of the Hurst exponent ranges between 0 and 1. A value of 0.5 indicates a true random walk (a Brownian time series). In a random walk there is no correlation between any element and future element. If the Hurst exponent is 0.5<H<1.0, the process will be a long memory process. Data sets like this are sometimes referred to as fractional Brownian motion. A Hurst exponent value in this range indicates persistent behaviour (or, a positive autocorrelation). A Hurst exponent value 0<H<0.5 will exist for a time series with anti-persistent behaviour (negative autocorrelation).

Astrophysical processes are highly nonlinear over wide ranges of temporal and spatial scales. Although the multiplicity of scales have been often considered and discussed, there is no consensus on the best way to deal with multi-scale phenomena. However, during the last two decades there has been a mushrooming interest in the application of scaling properties of Astrophysics, as an impetus for the development of new concepts, notions, formalisms and techniques [1].

Solar neutrino flux detection is very important not only to understand the stellar evolution but also to understand the origin of the solar activity cycle. Recent solar neutrino flux observed by Super-Kamiokande [2] and SNO detectors [3] suggest that solar neutrino flux from $^8$B neutrino and $^3$He+p neutrino from Standard Solar Model (S.S.M.)[4] is at best compatible with S.S.M. calculation if we consider the neutrino oscillation of M.S.W.[5] or if the neutrino flux from the sun is a mixture of two kinds of neutrino i.e. $\nu_e$ and $\nu_\mu$ [6]. Standard Solar Model (S.S.M.) are known to yield the stellar structure to a very good degree of precision but the S.S.M. cannot explain the solar activity cycle, the reason being that this S.S.M. does not include temperature and magnetic variability of the solar core [7,8]. The temperature variability implied a variation of the energy source and from that source of energy magnetic field can be generated which also imply a magnetic variability [8]. The temperature variation is important for the time variation of the solar neutrino flux. So we need a perturbed solar model and it is outlined by Raychaudhuri since 1971[7, 8], which may satisfy all the requirements of solar activity cycle with S.S.M.. For the support of perturbed solar model we have demonstrated that solar neutrino flux data are fractal in nature [9, 10, 11]. The excess nuclear energy from the perturbed nature of the solar model transforms into magnetic energy, gravitational energy and thermal energy etc. below the tachocline. The variable nature of magnetic energy induces dynamic action for the generation of solar magnetic field.

Solar neutrino flux data from Homestake [12] detector varies with the solar activity cycle but at present it appears that there is no significant anti correlation of solar neutrino flux data with the sunspot numbers. Many authors analysed the solar neutrino flux data from Homestake detector and have found short-time periodicities around 5, 10, 15, 20, 25 months etc.. Raychaudhuri analysed the solar neutrino flux data from SAGE, GALLEX, Superkamiokande and have found that solar neutrino flux data varies with the solar activity cycle and have found periodicities around 5 and 10 months.

The purpose of the present paper is to see whether the SAGE, GALLEX-GNO solar neutrino flux data are variable in nature or not. The observation of a variable nature of solar neutrino would provide significance to our understanding of solar internal dynamics and probably to the requirement of the modification of the Standard Solar Model i.e. a perturbed solar model.

**FINITE VARIANCE SCALING METHOD:**

A well known version of Finite Variance Scaling Method (FVSM) is the Standard Deviation Analysis(SDA) [13], which is based on the evaluation of the standard deviation $D(t)$ of the variable $x(t)$.

In a time series containing n data it yields [13, 14]

$$D(t_j) = [1/j \sum_{i=1}^{j} x^2(t_i) - \{1/j \sum_{i=1}^{j} x(t_i)\}^2]^{1/2} \qquad (1)$$

for $j = 1,2,\ldots,n$

Eventually it is observed [13,14]

$$D(t) \, \acute{\alpha} \, t^H \qquad (2)$$

The exponent H is interpreted as the Hurst exponent. It is evaluated from the gradient of the best fitted straight line in the plot of log $\{D(t)\}$ against log t.

The relationship between the fractal dimension D and the Hurst exponent H is

$$D = 2 - H. \qquad (3)$$

**RESULTS AND DISCUSSION:**

We have applied the FVSM on the monthly solar neutrino flux data from (1) SAGE detector during the period from 1$^{st}$ January 1990 to 31$^{st}$ December 2000; (2) SAGE detector during the period from April 1998 to December 2001; (3) GALLEX detector during the period from May 1991 to January 1997; (4) GNO detector during the period from May 1998 to December 2001; (5) GALLEX-GNO detector (combined data) from

May 1991 to December 2001 and (6) average of the data from GNO and SAGE detectors during the period from May 1998 to December 2001. Figures 1-6 show the plots of D (t) against t calculated from the above-mentioned data. These plots are fitted with equation (2) and (3) to yield H and D for different solar neutrino flux data in the following way:

| Experiment | Hurst Exponent(H) | Fractal Dimension(D) |
|---|---|---|
| (1)SAGE detector during the period from $1^{st}$ January 1990 to $31^{st}$ December 2000 | .071 | 1.929 |
| (2) SAGE detector during the period from April 1998 to December 2001 | .519 | 1.481 |
| (3) GALLEX detector during the period from May 1991 to January 1997 | .111 | 1.889 |
| (4) GNO detector during the period from May 1998 to December 2001 | .268 | 1.732 |
| (5) GALLEX-GNO detector (combined data) from May 1991 to December 2001 | .108 | 1.892 |
| (6)Average of the data from GNO and SAGE detectors during the period from May 1998 to December 2001 | .353 | 1.647 |

From the above table it is clear that the values of H are below than 0.5 for the monthly solar neutrino flux data in 1), 3), 4), 5) and 6) which in turn suggest that they exhibit anti-persistent behaviour (or, negative autocorrelation) and has underlying trends so that we can conclude that these data are periodic in nature while the monthly solar neutrino flux data in 2) shows the tendency of a long memory process as in this case the value of H is obtained to be slightly greater than 0.5.

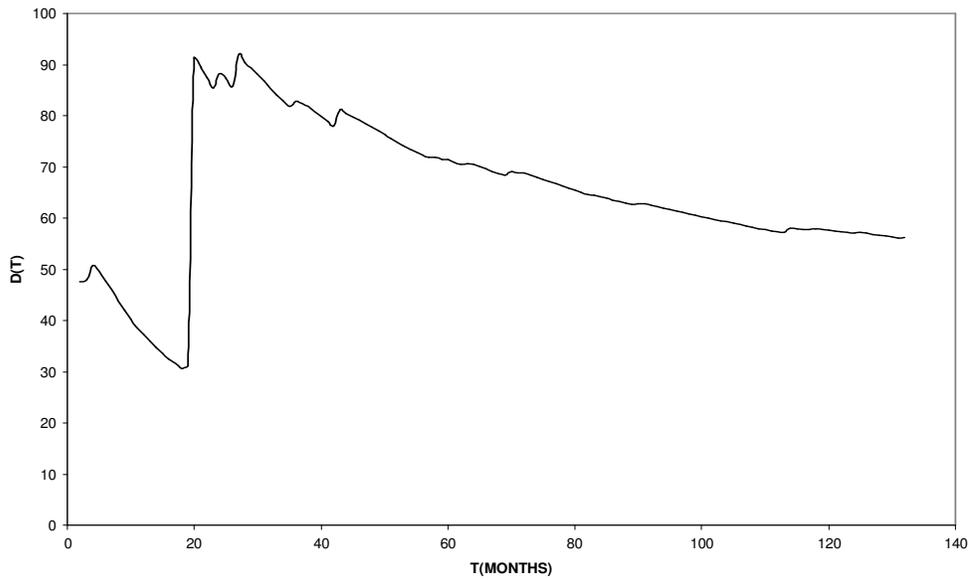

FIG.1: ANALYSIS OF SAGE DATA BY FINITE VARIANCE SCALING METHOD

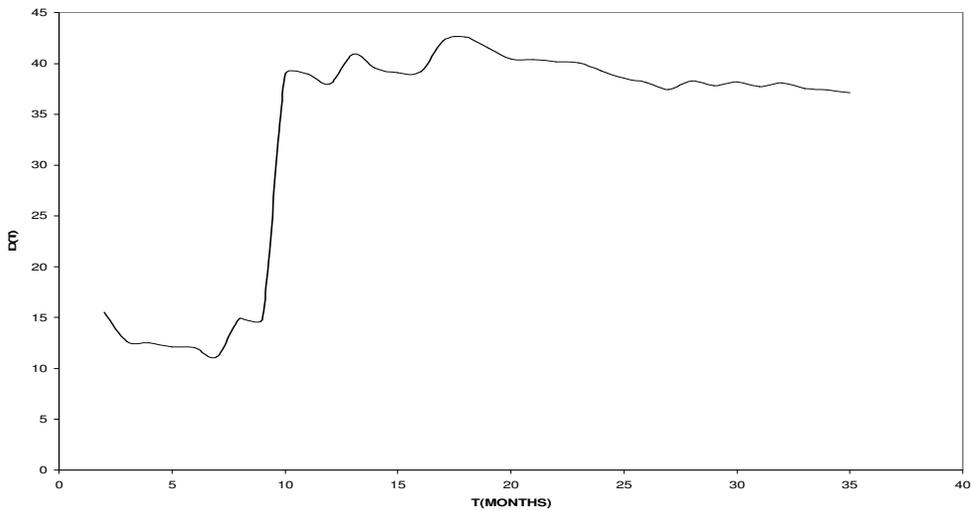

FIG.2: ANALYSIS OF SAGE DATA BY FINITE VARIANCE SCALING METHOD

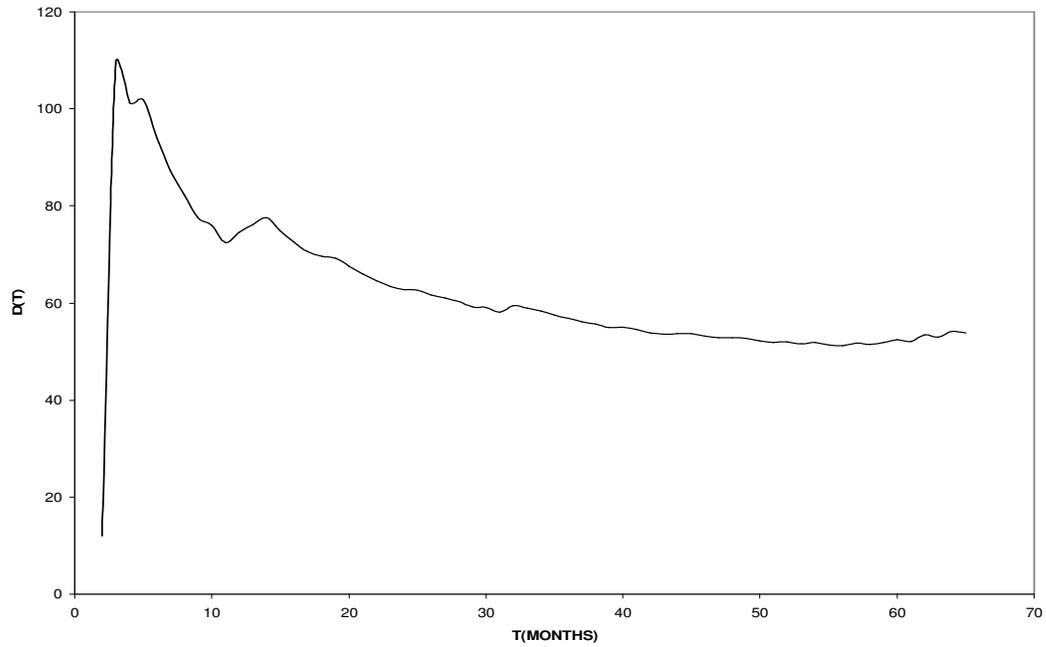

**FIG.3: ANALYSIS OF GALLEX DATA BY FINITE VARIANCE SCALING METHOD**

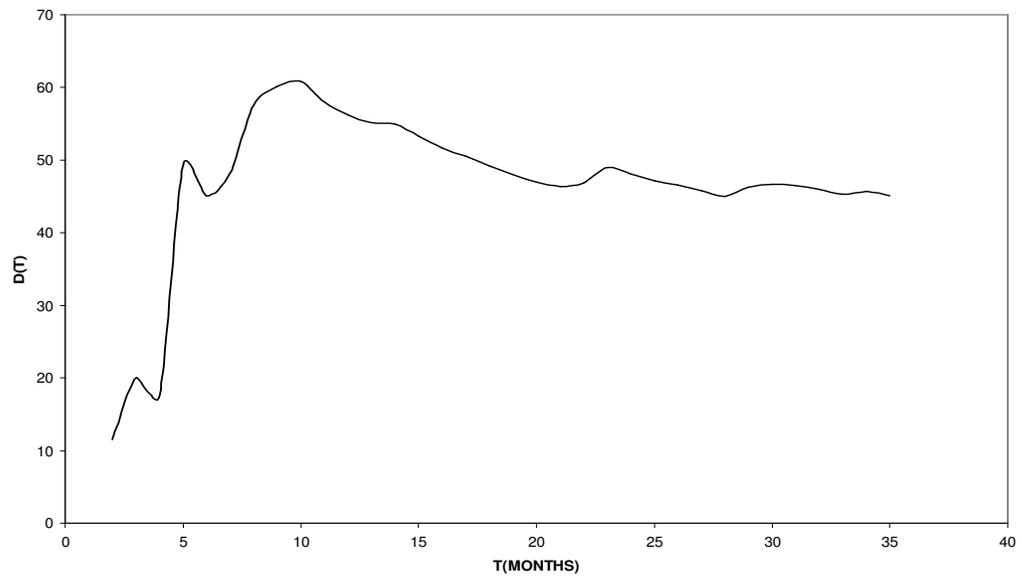

**FIG.4: ANALYSIS OF GNO DATA BY FINITE VARIANCE SCALING METHOD**

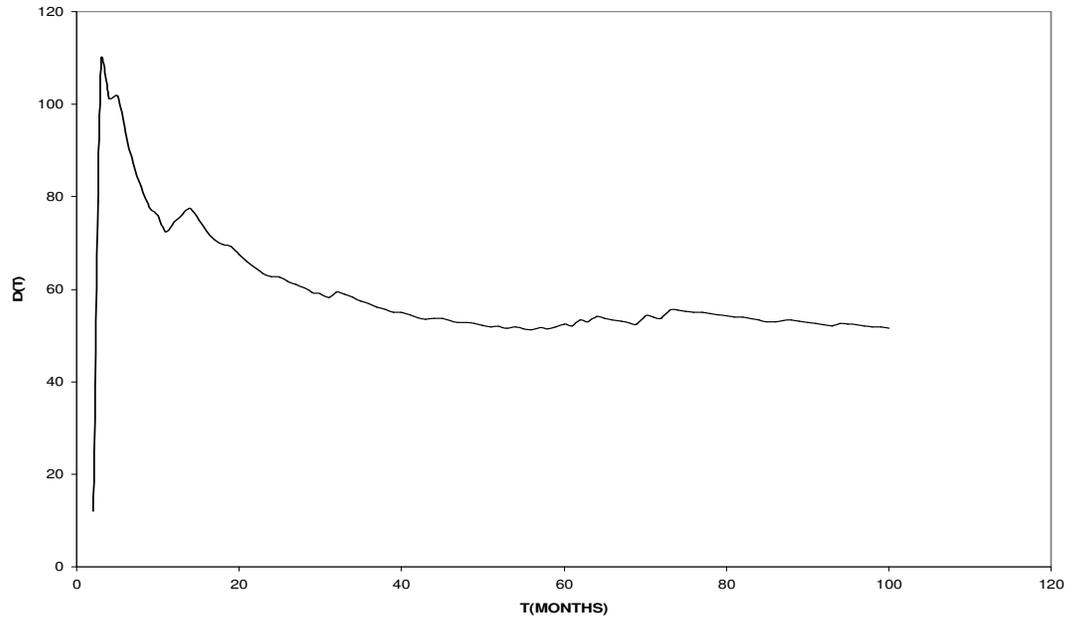

**FIG.5: SCALING ANALYSIS OF COMBINED GALLEX-GNO DATA BY FINITE VARIANCE SCALING ANALYSIS**

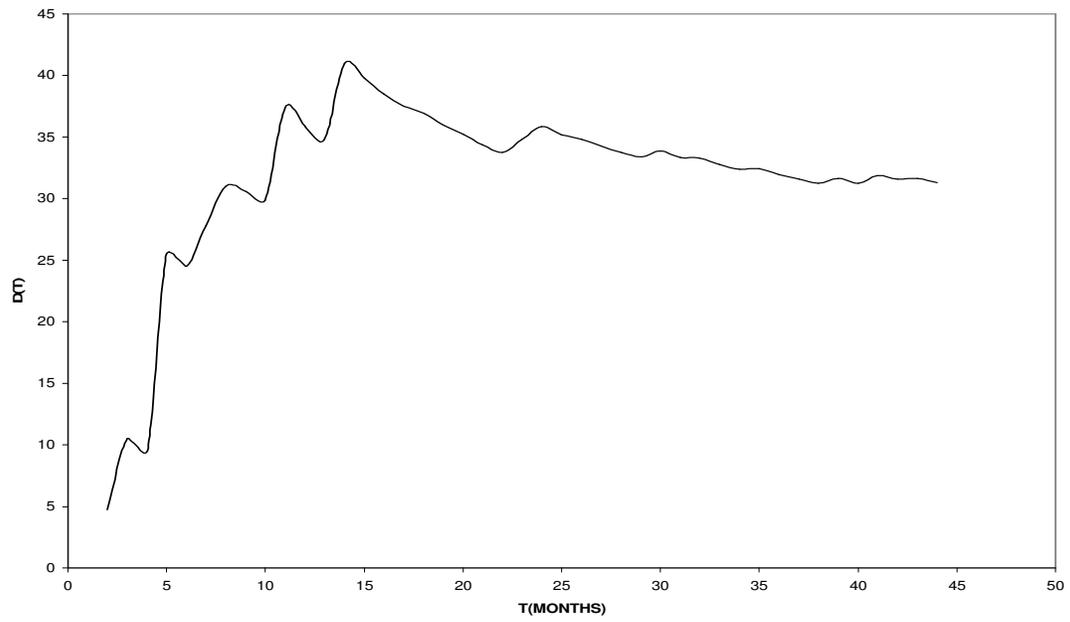

**FIG.6: ANALYSIS OF THE AVERAGE OF GNO AND SAGE DATA BY FINITE VARIANCE SCALING METHOD**